# The Role of Schema Matching in Large Enterprises

Ken Smith, Peter Mork, Len Seligman, Arnon Rosenthal, Michael Morse, Christopher Wolf,
David Allen & Maya Li
The MITRE Corporation
7515 Colshire Dr
McLean, VA 22102
(703) 983-6115

{kps, pmork, seligman, arnie, mdmorse, cwolf, dmallen, haoli}@mitre.org

## ABSTRACT
To date, the principal use case for schema matching research has been as a precursor for code generation, i.e., constructing mappings between schema elements with the end goal of data transfer. In this paper, we argue that schema matching plays valuable roles independent of mapping construction, especially as schemata grow to industrial scales. Specifically, in large enterprises *human decision makers and planners* are often the immediate consumer of information derived from schema matchers, instead of schema mapping tools. We list a set of real application areas illustrating this role for schema matching, and then present our experiences tackling a customer problem in one of these areas. We describe the matcher used, where the tool was effective, where it fell short, and our lessons learned about how well current schema matching technology is suited for use in large enterprises. Finally, we suggest a new agenda for schema matching research based on these experiences.

## Categories and Subject Descriptors
H.2.1 [**Database Management**]: Logical Design – *data models, schema and subschema.*

## 1. Introduction
The database community has been conducting research on schema matching for decades [1]. This research has usually assumed that schema matching (i.e., the generation of semantic correspondences among schemata) is merely a precursor to the creation of executable transformation code (e.g., database views or ETL scripts). Thus, the assumption has always been that the ultimate result of schema integration will be a program to effect data transfer across systems with different yet similar schemata. For example, Bernstein, et al. state that "Most systems-integration work requires creating mappings between models, such as database schemata, message formats, interface definitions, and user-interface forms [2]." Indeed, when we embarked on our own Harmony schema matching project [3], we believed that to fully realize its value, we would need to integrate Harmony with a schema mapping tool, and we did so with the AquaLogic Data Service Platform [4, 5].

However, while working on Harmony, members of our research team have worked closely with customers in large government organizations on a wide range of integration problems. In the course of those efforts, we came to the realization that *schema matching provides crucial insights to human decision makers,* including planners, information system managers, and enterprise architects who have no immediate intention of performing data exchange.

We also observed that the current generation of schema matching tools lacks capabilities that would help provide these necessary insights; for example, enterprise architects and managers often work with much larger schemata than developers need to manipulate. In addition, the challenges of supporting architects, decision makers, and planners in large enterprises (e.g., corporations, government organizations) introduce *new* requirements for schema matching research and tools.

Based on our insights from these experiences, this paper makes the following contributions:

1. We describe several application areas that illustrate the use of schema matching results by human decision makers, planners, and system architects in large enterprises. (Section 2).

2. For one such application area—a planning effort involving multiple large schemata (each on the order of $10^3$ elements)—we report our experiences and describe our lessons learned. (Sections 3 and 4).

3. Motivated by these observations and experiences, in Section 5 we then enumerate additional research challenges that must be addressed to meet needs of data integrators in large enterprises.

## 2. Broadening the view of schema matching

In this section, we describe several scenarios drawn from our work with government customers in which the results of schema matching are directly useful to people and organizations, without ever generating transformation code.

**Project feasibility:** Within the US Department of Defense (DoD), many integration projects are organized around "communities of interest" (COIs) [6]. Each such community represents a collection of parties whose information needs overlap significantly and who have in interest in sharing data. A key goal of a COI is to develop a *community vocabulary* (i.e., a mediated schema) that can serve as a basis for data exchange.

Convening a COI requires a decision maker to commit significant resources to support the COI's activities; however, no resources will be committed unless the potential value is clear. Schema matching tools are needed to quickly estimate the extent to which it will be feasible to generate a community vocabulary from a collection of data sources.

**Project planning:** A related scenario involves determining the level of effort (and corresponding cost) needed for an integration project. For example, the Air Operations COI has presented us with a community vocabulary and asked: To what extent can the attributes in the community vocabulary be populated by a specific data source? To what extent must this vocabulary be expanded to cover the data used by a related set of systems? And, most importantly, how much time and money should be allocated to these projects?

Answering these questions requires us to match the source schemata to the community vocabulary, although not to generate mappings (at this point), but to help the COI planners estimate the level of programming effort required to establish the actual mappings so an appropriate contract can be written with realistic cost estimates.

**Generating an exchange schema:** In the preceding scenario, the COI has already identified a candidate vocabulary. However, it is often the case that COI participants need to construct an exchange schema from scratch. In many scenarios, this exchange schema needs to be constructed quickly, albeit incompletely. For example, in an emergency response scenario, many new data sharing partners (e.g., state and federal agencies, non-profits, corporations) may suddenly be thrust together to respond to a crisis.

In such a scenario, schema matching tools are needed to identify which information the available systems have in common. The various agencies need to be able to throw their data models into a giant beaker and to distill out a *minimal* mediated schema that will serve as the basis for their collaboration. Eventually, the agencies will need to map their systems to the mediated schema, but they need schema matching support while at the negotiating table.

**Identifying the integration target:** In certain large and diverse communities (e.g., the entire U.S. military, the entire U.S. law enforcement community), authorities mandate particular data models as a starting point for much data sharing. Under the pressure of accommodating a broad range of possible data exchanges, these schemata can grow to become too large for participants to comprehend and exploit.[1] In this situation, data sharing partners need schema matching support to identify that subset of the exchange schema that is relevant to their system. Again, schema matching provides valuable information at the project design stage, well before programmers are tasked with developing mappings.

**Enterprise information asset awareness:** The Chief Information Officer (CIO) of a large enterprise needs to understand what information is being managed across the enterprise's information systems, and by which systems. For example, the CIO of a Health Maintenance Organization may want to know which data sources contain the concept of "blood test", and what specific concepts are shared by five different systems involving patient health history.

The CIO could answer these questions given a *comprehensive vocabulary*: an exhaustive list of the concepts found in a set of data sources, and, for each concept, the sources using that concept in their data model. Schema matching is a valuable tool for forming a comprehensive vocabulary because it helps identify shared concepts, independently of future mapping construction. In Section 3 we describe an effort to derive a comprehensive vocabulary in support of project planning.

**Finding relevant and related schemata:** Some very large communities have an enterprise schema registry. For example, within the DoD, all data models must be posted to the DoD Metadata Registry (MDR), which now contains thousands of schemata in a wide range of formats. One goal of the MDR is to make system developers more aware of each other's systems to increase reuse and thereby reduce semantic heterogeneity.

A powerful way to search the MDR would be to simply use one's target schema as the "query term." Using schema matching technology, the system would rank the available schemata.

Using a similar approach, a schema repository such as the MDR could automatically propose new COIs by clustering the schemata into related groups. Automatic schema clustering within a repository would also be valuable to an enterprise's CIO, who seeks to understand the extent to which large and diverse sets of information assets across their organization are related.

## 3. Detailed Example

In this section we present an instance of the project planning application area in some detail.

### 3.1 Task Summary

A military customer presented us with two large independently-developed schemata that the customer believed should overlap significantly. Each schema should, for example, include information about persons, vehicles, and military units.

- Schema A ($S_A$) is relational, contains 1378 elements, and is the schema of version 3 of an actively used system: $Sys(S_A)$. To enhance compatibility among military services, $Sys(S_A)$ is currently being redesigned into version 4.
- Schema B ($S_B$) is an XML Schema, contains 784 elements, is the schema of legacy system $Sys(S_B)$, and is reputed by the customer to include a conceptual subset of $S_A$. $Sys(S_B)$ is disliked by users and there is pressure to eliminate it.

The customer recognized the transition from $Sys(S_A).v3$ to $Sys(S_A).v4$ as an opportunity to eliminate $Sys(S_B)$ by augmenting $Sys(S_A)$ to subsume its role, incorporating its operations and any distinct data elements of $S_B$ into $S_A$. However, the customer lacked detailed expertise with $Sys(S_B)$, and the specific nature and extent of the relationship between $S_A$ and $S_B$ was unknown.

---

[1] E.g., en.wikipedia.org/wiki/JC3IEDM and niem.org

Eliminating Sys($S_B$) was *not* the clear choice if a) the set of distinct $S_B$ elements were sufficiently large and b) the set of common elements (i.e., elements involved in matches between $S_A$ and $S_B$) were sufficiently small. In this case, it could be inferred that the roles of Sys($S_A$) and Sys($S_B$) were so different that subsuming Sys($S_B$) was unattractive compared to retaining it and building an ETL bridge to feed Sys($S_A$) (i.e., a classic data warehouse architecture). In summary: due to the size, complexity, and unknown relationship of these systems, the customer could not easily evaluate the merit and cost of these two options.

As an initial step toward selecting one of these two options, the customer presented us with schemata $S_A$ and $S_B$ and requested an analysis of what they held in common, how and to what extent they differed, and a representation of their *comprehensive vocabulary*. This customer challenge appeared to be an excellent test of schema matching technology's ability to address a real, "industrial scale" problem.

Note several important features of this problem:

1. Both the *distinctiveness* (i.e., $S_B$–$S_A$ and $S_A$–$S_B$) and the *overlap* (i.e., $S_A \cap S_B$) of the data models involved were important. In particular, the cardinalities of $S_A \cap S_B$ and $S_B$–$S_A$ were vital to the customer's decision process.
2. The customer organization was so large that the people tasked with planning the fate of information systems lacked detailed knowledge of those systems' contents.
3. The scale of the entailed schema match, $10^6$ potential matches, would be tedious for human users, and exceeds that of most published schema matching studies.
4. Generating executable mappings between $S_A$ and $S_B$ was *not* the immediate goal.

## 3.2  Harmony Schema Matcher

We now describe the matcher used to address this problem. Harmony is a MITRE-developed schema matcher that combines multiple match algorithms with a graphical user interface for viewing and modifying the identified schema correspondences. Unlike most schema matching tools, Harmony relies heavily on textual documentation to identify candidate correspondences instead of data instances because, as we reported in [3], at least in the government sector, schema documentation is easier to obtain than data (which may not yet exist, or may be sensitive). This characterization holds for the matching task between $S_A$ and $S_B$.

The Harmony match engine adopts a conventional schema matching architecture [2, 7-9]. It begins with linguistic preprocessing (e.g., tokenization and stemming) of element names and any associated documentation. Then, several match voters are invoked, each of which identifies correspondences using a different strategy. For each [source element, target element] pair, each match voter establishes a confidence score in the range (–1, +1) where –1 indicates that there is definitely no correspondence, +1 indicates a definite correspondence and 0 indicates complete uncertainty. A vote merger combines the confidence scores into a single match score.

As a match voter observes more evidence, the confidence score is pushed towards –1 or +1. Compared to conventional schema matching tools, Harmony is novel in that it considers both the standard evidence ratio (e.g., number of shared words in the documentation) as well as the total amount of available evidence when calculating confidence scores. This approach allows the vote merger to combine confidence scores into a single match score based on how confident each match voter is regarding a given correspondence.

Once the match engine identifies candidate correspondences, the Harmony GUI displays these correspondences as color-coded lines between source and target elements. The GUI supports a variety of filters that then assist the integration engineer in focusing their attention in various ways. These filters are loosely categorized as *link filters*, which depend on the characteristics of a given candidate correspondence, and *node filters*, which depend on the characteristics of a given schema element.

For the current discussion, the most relevant link filter is the confidence filter. Only those correspondences whose match score falls within the specific range of values are displayed graphically. Thus, the integration engineer can focus their attention first on the most likely correspondences.

The node filters include a depth filter and a sub-tree filter. The former enables only those schema elements that appear at a particular nested depth. For example, in a relational model, relations appear at a depth of one and attributes at a depth of two. Using this filter, the engineer can focus on a particular level of granularity. The sub-tree filter enables only those elements that appear in a given sub-tree. For example, this filter can be used to focus one's attention on the 'Vehicle' sub-schema. The engineers responsible for matching $S_A$ to $S_B$ relied heavily on this filter.

## 3.3  Our Approach

On the surface, our task was "simply" to perform a 1378×784 schema match, report the pairs of matching elements, and report separately the unmatched elements from each schema. In fact, we had recently scaled Harmony to perform matches of this size, and the fully automated match executed in 10.2 seconds.

However, there were two problems with simply reporting the matcher's results. First, neither the matcher's output (a match matrix) nor existing visualizations of such a matrix gave our customer much insight into the high level areas of overlap and differences between $S_A$ and $S_B$. A lengthy list of matches such as "DATE_BEGIN_156 ⇔ DATETIME_FIRST_INFO" overloaded the user and failed to provide the "big picture." The integration engineers recognized a need to introduce concept labels representing important domain concepts such as "Event" and "Person," which in turn could be assigned to schema elements.

Second, the integration engineers needed help navigating thousands of potential matches in order to validate and annotate them (e.g., with additional semantics such as is-a or part-of). Again, concept labels helped them organize this process.

For example, the "All_Event_Vitals" table of $S_A$ consisted of attributes corresponding to a concept they labeled "Event." In general, there was strong correlation between the tables and views in $S_A$, the types and elements in $S_B$, and these abstract designations. Through inspection, they identified 140 schema elements corresponding to useful abstract concepts in $S_A$ and 51 in $S_B$.

Once identified, they used Harmony's sub-tree filter to incrementally match each concept (i.e., the schema sub-tree rooted at that concept) with the entire opposing schema. For example, "All_Event_Vitals" in $S_A$ was chosen as the current sub-tree, and then matched to all of $S_B$. These match operations were rapid: typically between $10^4$ and $10^5$ matches were considered in each increment. Using the confidence filter, matches scoring above a threshold were then examined by a human integration engineer; valid matches and related annotations were recorded in Harmony. A common outcome was a strong match from the fields of one concept to the fields of a corresponding concept in the other schema, plus a few matching fields from other concepts. When this occurred, we also recorded a *concept-level match* (i.e., a match between a label used in $S_A$ and one used in $S_B$. 24 of these concept-level matches were thus identified and recorded.

This concept-at-a-time workflow had several benefits: It helped the integration engineers organize and track their progress each day. It also provided them with deeper insights into the conceptual coverage in each schema, and where these concepts matched. Finally, it allowed the integration engineers to keep entirely visible at least one side of the match (i.e., corresponding to one concept label), and perhaps both sides, in the user interface. This precluded a large mass of criss-crossing lines, denoting off-screen matches, from cluttering the display and obscuring the user's view.

The entire matching process required three days of effort, by two human integration engineers.

### 3.4 Outcome

At the customer's request, the final result was delivered as an Excel spreadsheet. The first sheet enumerated the 191 concepts with their 24 concept-level matches (167 rows), the second sheet contained the individual schema elements (indexed to a concept) and their element-level matches. Both sheets were organized in "outer-join" style with three types of rows: those specific to $S_A$, those specific to $S_B$, and those having matched elements of $S_A$ and $S_B$. Note that, had the end-goal been code generation, this step in the workflow would normally not have been needed.

The result showed that only 34% of $S_B$ matched $S_A$ and 66% of $S_B$ (or 517 elements) did not, indicating that subsuming Sys($S_B$) would be a challenging undertaking.

Near the completion of this project, the customer requested a further, expanded, study. They gave us four additional large schemata: $S_C$, $S_D$, $S_E$, and $S_F$, and requested a comprehensive vocabulary for $S_A$ and these four additional schemata. That is, for any non-empty subset of $\{S_A, S_C, S_D, S_E, S_F\}$, the customer wanted to know the terms those schemata (and no others in that group) held in common.

While this expansion is an ongoing effort, several clear lessons about the role of schema matching in large scale enterprises have emerged through our engagement in these tasks, as discussed in the following section.

### 4. Lessons Learned

In this section, we describe lessons learned from applying Harmony to a real problem. We begin by describing Harmony-specific features that proved to be useful. We then describe more general lessons about needed advances in schema match technology.

### 4.1 Harmony-specific Features

Three Harmony features were unexpectedly useful. 1) As mentioned above, the sub-tree filter was valuable. We could visually isolate a specific concept and its descendants, and limit a match operation to just those elements, ensuring the only match lines generated would originate in the selected sub-tree. Upon reflection, the sub-tree filter enables a form of *incremental schema matching*, a technique recommended for industrial scale problems [10]. 2) Similarly, the depth filter allows a user to ignore (i.e., exclude from matching) schema elements whose depth in the schema tree exceeds a certain threshold. This made it possible to only match table names in $S_A$, and ignore their attributes. 3) Finally, it was helpful that Harmony can export validated matches as a spreadsheet. Beyond simply being the desired delivery format, our integration engineers found it significantly easier to validate matches in this view, as opposed to the traditional line-drawing interface provided by Harmony and most schema matchers.

### 4.2 Lesson #1: Matching at Large Scales Requires Summarization

Large-scale schema matching involves a human workflow and not just running an algorithm. While some authors raise the workflow issue [3, 10], most of the schema matching literature focuses solely on the **MATCH ($S_1$, $S_2$)** [11] operator and its optimization. We observed that the integration engineers' workflow encompassed operations beyond the core task of matching. This workflow followed three basic steps:

1. **SUMMARIZE ($S_A$)** and **SUMMARIZE ($S_B$)**. In **SUMMARIZE(S)**, the human integrator creates a simpler version of S. This summarization allows the integrator to organize his cognitive efforts, may guide subsequent matching steps, and helps the integrator understand the final match product.
2. Using the schema matcher to do automated matching and interactive refinement of matches.
3. Post-matching analysis, in which both the matches and non-matching elements of $S_A$ and $S_B$ are exported to downstream tools for further analysis (more on this step in Lesson #2).

In the use case described above, neither our integration engineers, nor the customer could directly comprehend the result of **MATCH ($S_A$, $S_B$)**. Instead, the integration engineers manually summarized both $S_A$ and $S_B$ prior to performing the match. As noted above, they used a very simple summarization technique: creating a set of labels (corresponding to important domain concepts) and assigning them to particular schema elements.

This summarization served many purposes: First, both the customer and the integration engineers, who often lack detailed domain expertise (especially for legacy systems such as Sys($S_B$)), could more easily understand the conceptual coverage of each schema. Second, the integration engineers could first match the top-level nodes corresponding to those concepts, before diving into the lower-level details. Intuitively, one does not expect attributes from dissimilar concepts to match (although the integration engineers did observe some cross-concept matches). Third, the integration engineers could communicate their results to the customer at the level of the concepts (e.g., 75% of concept A matched, but only 25% of concept B matched).

Thus, we believe that industrial-scale schema matching systems must also support summarization. This operator would take a schema S as its input and generate a simpler representation S′ as its output. The operator must also generate a mapping that relates the elements of S to those of S′. While our engineers created S′ as a flat list of concept labels and the mapping related each schema element to at most one concept, recent research points the way toward more sophisticated schema summarization [12, 13]. More work on this is needed. In addition, it is an open question how to best use schema summarization to improve matching.

At a minimum, an industrial-strength schema matching tool should allow users to: a) manually summarize a schema, possibly by associating concepts with portions of the schema, b) match only those schema elements associated with a given element of the summarization, and c) easily iterate through the higher-level concepts to perform *incremental* matching, in the fashion described in [10].

## 4.3 Lesson #2: Matching at Large Scales Requires New Visualization Approaches

As noted by others [10], we found that "line-drawing" visualizations of schema match break down rapidly as schema size grows much larger than the user's screen. While this was ameliorated by Harmony's sub-tree filter, other visualization improvements would have made life much easier for our engineers.

First, we need a match-centric view of matches in addition to the typical schema-centric view. We came to this finding by probing the reasons that our customers wanted results delivered as a spreadsheet. At first, this struck us as a step backwards: most non–tool-assisted schema matching is done by manually editing a spreadsheet, and we (like others) viewed this approach as inferior to one based on a semi-automated schema matcher. Upon reflection, however, we found a problem with typical matcher interfaces: each schema remains intact while overlaid lines denote the matches. In many contexts, users care more about matches and sets of matches than about the original schema. Spreadsheets allow users to flexibly sort matches (e.g., by status, team member assigned to investigate it, etc.). This kind of match-centric view is something that must be added to schema match tools.

Second, the modified schema match workflow discussed above involves a summarization operation in which the integration engineers assigned domain concept labels to schema elements. It is an open question how to exploit these (and more sophisticated) summarizations to aid user understanding of matches and to best support the engineers' workflow.

## 4.4 Lesson #3: Schema-Matching Identifies both Commonalities and Distinctions

When matching is only performed as a precursor to code generation, it is natural to ignore unmatched elements. However, this use case demonstrates schema matching being used *to generate knowledge vital to planners and decision makers*. Both matched *and unmatched* elements can generate valuable knowledge. Specifically: we observed that the three sets: $\{S_1–S_2\}$, $\{S_2–S_1\}$, and $\{S_1 \cap S_2\}$ provide a useful partition of the match of two large schemata; the first two of these are examples of knowledge derived from unmatched schema elements.

We therefore assert that schema matching tools must provide as output not simply the set of full matches, but also any set of partial matches, including sets of unmatched elements. Note that summarization and visualization are again relevant: while it is trivial to produce $\{S_1–S_2\}$, $\{S_2–S_1\}$, and $\{S_1 \cap S_2\}$ from the output of a standard schema matcher, how does one give the integration engineer the big picture of what concepts are unique to one schema as opposed to another? This task becomes increasingly difficult as we address scaling beyond binary matches, the topic of the next subsection.

## 4.5 Lesson #4: Scaling Beyond Binary Matches is a High Priority

As we encountered in the recent expansion of our project in Section 3, addressing the needs of large enterprises can involve scaling *both* to large numbers of elements in individual schemata *and* to more than two schemata.

Consider again the three sets: $\{S_1–S_2\}$, $\{S_2–S_1\}$, and $\{S_1 \cap S_2\}$ that partition a binary match. In general, given N schemata there are $2^N-1$ such sets partitioning their N-way match; each of which supplies a potentially valuable piece of knowledge to information system decision makers, such as our customer.

Although current schema matching research focuses on binary matches, the challenge problems discussed in this paper require matchers that support matching multiple schemata (N>2) [1]. This requirement is especially important for scenarios involving communities of peer organizations who share data—e.g., recognizing likely sharing partners in the community, establishing information asset awareness (i.e., what info resources do we have as a community?), understanding which concepts would be most fruitful to try to standardize, etc.

## 5. Research Directions & Conclusions

In this section, we enumerate new research opportunities based on both our direct experiences and the use cases we identified.

**Schema summarization:** As noted above, our integration engineers needed to manually identify top-level concepts as a way to organize their work. We believe that schema summarization is a useful pre-cursor to large scale schema matching and that research is needed both in exploiting such summaries, and in creating them. Some promising work [12, 13] has been done, based on purely structural hints. More work is needed to extract key concepts from a schema and its documentation and to break the schema into semantically-related chunks.

Ideally, a summarization tool would convert a complex schema into a simpler representation, while preserving the relationship between the complex representation and the simple one. More theoretical work is needed to formalize this intuition, but the result of summarization should have the following characteristics: a) it is easier for a human to understand, b) it allows coarse-grained schema matching, c) it drives incremental refinement, and d) it helps the user understand the final match result. We expect there to be both algorithmic and user interface components of this work.

**User interfaces:** In our work and that of others [10, 14], there has been new emphasis on the schema matcher user interface. The canonical UI represents the source and target schema as hierarchical structures with lines drawn between them.

We found this view of matching to be useful at times; however, the sheer number of lines displayed at once is often overwhelming. As a result, research prototypes such as [3] and [10] provide mechanisms for reducing the number of lines shown at any one time. In addition to de-cluttering, our engineers needed a match-centric view that would let them flexibly sort and group matches. In addition, future engineers will need interfaces appropriate for specifying schema summarizations and visualizing their results. These recommendations are just a first step. Much more UI work is needed, preferably using rigorous HCI experimental methods.

**N-way matching**: Strategies for matching more than two schemata are not novel; the first general discussion of which we are aware was over two decades ago [1]. However, it has only been infrequently considered since that time, e.g., [15, 17], and this is a research area which needs a revival. Deriving a comprehensive vocabulary for a group of schemata in a large enterprise relies on extending schema matching beyond the binary case. The larger-N use cases we have presented also lead to more research problems than simply matching, as illustrated by the following three topics.

**Schema clustering and overlap analysis**: Some use cases in Section 2 identify, at a high-level, the overlap between two (or more) schemata. We need new techniques to characterize overlap approximately but quickly, in a way that is meaningful for human decision makers.

Numeric characterizations of overlap could also be used as inter-schema distance metrics by a clustering algorithm. The ability to identify clusters of related schemata is vital, providing CIOs with a big picture view of enterprise data sources and revealing to integration planners the most promising (i.e., tightly clustered) candidates for integration. Schema clustering techniques have been presented for XML DTD's [16] and conceptual schemata (e.g., ER diagrams) [13], and a plan to topically cluster schemata in the PAYGO architecture is discussed in [18], however much more research is needed, both into algorithms for generating schema clusters as well for appropriate means to visualize them.

**Schema search:** Schema matching tools are used after integration engineers identify the schemata to be matched. Complementary search tools are needed to *locate* potential match candidates from a larger pool of schemata. These would take, as input, a query specification (e.g., an example schema, predicates over schema characteristics, example instance values). A simple search tool would return a list of schemata sorted by relevance to the query; a more sophisticated one could return relevant schema fragments.

**Enterprise metadata repositories:** Large enterprises can have hundreds to thousands of schemata, illustrating the need to manage schemata as data themselves. A schema (metadata) repository is an appropriate context in which to cluster schemata, to summarize them, to search for match candidates and to store resulting match information.

Several commercial repository tools are available, but these ignore the importance of schema matches as knowledge artifacts. However, as noted in [7, 18], other developers should be able to benefit from previous matches. Fundamental to such a repository is the notion that matches are context-dependent; a match that supports search may not have sufficient precision to support a business intelligence application. A related research topic is managing matching provenance—i.e., who said that X is the same as Y, and should I trust that assertion in *my* application?

**Support for integration teams**: As illustrated by our experiences in Section 3, large-scale schema matching is rarely performed by a single individual with domain expertise in all the relevant schemata, as well as knowledge about data integration. Research is needed to enhance the current generation of schema matchers with support for *integration teams*, with members having different sorts of expertise. For example, how can we divide very large matching workflows into modular task queues appropriate to each team member, along with the necessary communication mechanisms, to support a team-based matching effort? The appropriate task visualization may also vary, for team members with different expertise.

**CONCLUSION**

The push toward realistic, industrial-scaled schema matching problems changes the problem both quantitatively and qualitatively. Algorithmic improvements for binary matching are valuable, but just one requirement. In large enterprises involving many information systems, we observe that human planners and decision makers can benefit as primary consumers of the information generated by schema matching, as opposed to these results solely being "piped" into code generation. Human consumers, however, require different products from a schema match. Our experiences thus point to a broader agenda for schema matching research, opening new and interesting areas for exploration, such as: a tighter integration of schema matching and schema summarization research, specialized user interfaces, and support for larger-*N* schema operations such as clustering, search, and repositories.